\documentclass[aps,preprint,amsmath,amssymb,showpacs]{revtex4}
\usepackage{graphicx}
\def\be{\begin{eqnarray}}
\def\en{\end{eqnarray}}
\def\non{\nonumber}

\def\pl{{ Phys. Lett.}~}
\def\pr{{ Phys. Rev.}~}
\def\prl{{ Phys. Rev. Lett.}~}
\def\bi{\bibitem}
\begin{document}

\title{\Large \bf Isospin mass splittings of heavy baryons in HQS
 }

\author{ \bf \large Chien-Wen Hwang\footnote{Email:
t2732@nknucc.nknu.edu.tw} and Ching-Ho Chung\footnote{Email:
chung.cigo@gmail.com}
 }

\affiliation{\centerline{Department of Physics, National Kaohsiung Normal University,} \\
\centerline{Kaohsiung, Taiwan 824, Republic of China}
 }


\begin{abstract}
In this paper, the electromagnetic mass differences of heavy hadrons
are discussed, while ignoring the relevant hyperfine interactions.
The effects of one-photon exchange interaction and up-down quark
mass difference are parameterized. Two mass difference equations
$2\Sigma_c^+ - (\Sigma_c^{++} + \Sigma_c^0) = 2\Sigma_b^0 -
(\Sigma_b^+ + \Sigma_b^-)$ and $(\Xi_{cc}^+ - \Xi_{cc}^{++}) +
(\Xi_{bb}^- - \Xi_{bb}^{0}) =2(\Xi_{bc}^0 - \Xi_{bc}^{+})$ for the
heavy baryons are obtained. In addition, the masses of $\Sigma_b^0$,
$\Xi_b^0$, and $\Xi_{cc}^{++}$ are predicted based on the known
experimental data.
\end{abstract}
\maketitle %

\section{Introduction}
For the heavy baryons which contain one heavy quark, all the
$s$-wave charmed sector have been found at present. However, except
for particle $\Lambda_b^0$ which was found in the early 1980's,
there has been no significant progress in searching $s$-wave
bottomed sector until last year. Recently, some bottomed baryons
were discovered at Fermilab. They are the exotic relatives of the
proton and neutron $\Sigma^{(*)+}_b$ and $\Sigma^{(*)-}_b$ by CDF
collaboration \cite{CDF} and the triple-scoop baryon $\Xi_b^-$ by D0
and CDF collaborations \cite{DZero,CDF1}. In addition, for the heavy
baryons which contain two heavy quarks, only the doubly charmed
baryon $\Xi_{cc}^+$ has been observed by SELEX collaboration
\cite{SELEX1,SELEX2} (in fact, the BABAR \cite{Babar} and BELLE
\cite{Belle} experiments failed to observe the SELEX states). It is
reasonable that the remainder of $s$-wave heavy baryons, which
include (i) the double strange baryons $\Omega_b$ and doubly heavy
baryons $\Xi_{bc}$ and $\Xi_{bb}$, (ii) the excited states of, for
example, the triple-scoop baryon $\Xi'_b$ and $\Xi_b^*$, and (iii)
the isospin partners of the known baryons (namely, $\Sigma^0_b$,
$\Xi_b^0$, and $\Xi_{cc}^{++}$), will be observed in the foreseeable
future. The particles (i) and (ii) have been studied in some
researches. This paper focuses on the type (iii) particles, which
are based on the heavy quark symmetry (HQS).

Isospin or $SU(2)$ symmetry originates from treating the up and down
quarks as an isospin doublet. This symmetry is broken by the up-down
quark mass difference, and also by electromagnetic interactions,
which distinguish the different charges carried by the up and down
quarks. For the former contribution, the $u$ and $d$ quarks are
intrinsically light, and their bare mass difference is about several
MeV \cite{PDG08}. However, within the limit of a hadron, the $u$ and
$d$ quark masses are suitably described by the constituent values
which are about $350$ MeV greater than the intrinsic ones. In fact,
the precise values not only depend on the binding energies of
various quarks, but also on the context. Therefore, the effective
mass difference of $u$ and $d$ quarks is quite uncertain. In this
study, the detailed dynamics was not included, but was parameterized
the following evaluation. For the latter contribution, it is widely
accepted that quantum electrodynamics (QED) is the correct theory
for electromagnetic (EM) interactions. In QED, the photons mediate
EM forces among charged particles. Therefore, this paper aims to
discuss the one-photon exchange interaction between the different
quarks. As mentioned in Ref. \cite{Rosner}, the EM interaction
between $i$ and $j$ quarks leads to two kinds of energy
contribution. One is the Coulomb energy
 \be
 \Delta E_{coul}=\alpha e_i e_j \langle{\frac{1}{r_{ij}}}\rangle,
 \label{coul}
 \en
where $\alpha$ is the fine structure constant, $e_i$ is the charge
of quark $i$ and $\langle 1/r_{ij}\rangle$ is the expectation value
of the inverse distance between $i$ and $j$ quarks. In the flavor
$SU(3)$ limit, $\langle 1/r_{ij}\rangle$ is universal throughout a
multiplet. Another energy contribution is the EM hyperfine splitting
 \be
 \Delta E_{hf}^e=\textrm{const}\times \alpha e_i e_j |\Psi_{ij}(0)|^2
 \frac{\langle \sigma_i \cdot \sigma_j \rangle}{m_i m_j} \label{Ehf}
 \en
where $|\Psi_{ij}(0)|^2$ is the square of the $s$-wave function of
two quarks at zero relative separation and $\sigma_i (m_i)$ is the
spin (mass) of quark $i$. According to the conclusion of Ref.
\cite{Rosner} and the experimental data \cite{CDF,PDG08}, this EM
hyperfine splitting contribute to systematic uncertainty of the
experimental results and can be ignored if one of the quarks is
heavy.

The remainder of this paper is organized as follows. Section II
presents a brief review on the heavy quark effective theory (HQET).
Section III is the analyses of the heavy mesons and the heavy
baryons which contain one or two heavy quarks. Finally, the
conclusions are given in Section IV.

\section{Heavy quark effective theory}
It was found in 1989 that, within the limit $m_Q \to \infty$,
quark-gluon dynamics is independent of the heavy quark flavor and
spin \cite{iw}. This is called HQS, which is not present in the full
QCD Lagrangian. Thus, HQS is valid only when the typical gluon
momenta are much less than the heavy quark mass $m_Q$.

The full QCD Lagrangian for a heavy quark ($c$, $b$, or $t$) is
given by
 \be
   {\cal L}_Q = \bar Q~(i\gamma_\mu D^\mu - m_Q)~Q,   \label{Lag}
 \en
where $D^\mu \equiv \partial ^\mu - i g_s T^a A^{a\mu}$ with $T^a =
\lambda^a/2$. Inside a hadronic bound state containing a heavy
quark, the heavy quark $Q$ interacts with the light degrees of
freedom by exchanging gluons with the momenta of order
$\Lambda_{QCD}$, which is much smaller than its mass $m_Q$.
Consequently, the heavy quark is close to its mass shell, and its
velocity does not deviate much from the hadron's four-velocity $v$.
In other words, the heavy quark's momentum $p_Q$ is close to the
``kinetic" momentum $m_Q v$ resulting from the hadron's motion
 \be
   p^\mu_Q = m_Q v^\mu + k^\mu,  \label{Pk}
 \en
where $k^\mu$ is the so-called ``residual" momentum and is of order
$\Lambda_{QCD}$ and the corresponding change in the heavy quark
velocity vanishes as $\Lambda_{QCD}/m_Q \rightarrow 0$. Thus it is
appropriate to introduce the ``large" and ``small" component fields
$h_v$ and $H_v$ by
 \be
   h_v (x) = e^{im_Q v\cdot x} {\cal P}_+ Q(x),  \non \\
   H_v (x) = e^{im_Q v\cdot x} {\cal P}_- Q(x),
 \en
where ${\cal P}_\pm $ are the positive and negative energy
projection operators
 \be
   {\cal P}_\pm = \frac{1 \pm \not\!v}{2},
 \en
with ${\cal P}^2_\pm = {\cal P}_\pm$ and ${\cal P}_\pm {\cal P}_\mp
= 0$, and ${\cal P}_+$ satisfies the useful identity
 \be
   {\cal P}_+~\gamma^\mu~{\cal P}_+ = {\cal P}_+~v^\mu~{\cal P}_+.  \label{PP}
 \en
$h_v(x)$ and $H_v(x)$ are related to the original field $Q(x)$ by
 \be
   Q (x) = e^{-im_Q v \cdot x} \left[h_v(x) + H_v (x) \right].  \label{hH}
 \en
It is clear that $h_v$ annihilates a heavy quark with velocity $v$,
while $H_v$ creates a heavy antiquark with velocity $v$. In the
heavy meson's rest frame $v=(1,\vec 0)$, $h_v(H_v)$ correspond to
the upper (lower) two components of $Q (x)$. In terms of the new
fields, the QCD Lagrangian for a heavy quark given by (\ref{Lag})
takes the following form
 \be
   {\cal L}_Q = \bar h_v i v\cdot D h_v - \bar H_v (i v\cdot D + 2 m_Q) H_v +
   \bar h_v i\not\!\!D_{\bot} H_v + \bar H_v i\not\!\!D_{\bot} h_v  \label{nLag}
 \en
where $D^\mu_{\bot} = D^\mu - v^\mu v \cdot D$ is orthogonal to the
heavy quark velocity, $v\cdot D_{\bot} = 0$. In (\ref{nLag}), $h_v$
describes the massless degrees of freedom, whereas $H_v$ corresponds
to fluctuations with twice the heavy quark mass. The heavy degrees
of freedom represented by $H_v$ can be eliminated using the
equations of motion of QCD. By substituting (\ref{hH}) into $
(i\not\!\!D - m_Q) Q (x) = 0$ and multiplying it by ${\cal P}_\pm$,
we can obtain
 \be
   -i v \cdot D h_v = i \not\!\!D_\bot H_v,  \label{1-14}  \\
   (i v \cdot D + 2 m_Q) H_v = i \not\!\!D_\bot h_v. \label{1-15}
 \en
$H_v(x)$ can be eliminated to obtain the equation of motion for
$h_v$. It is easy to check that the resulting equation follows from
the effective Lagrangian
 \be
   {\cal L}_{Q,eff} = \bar h_v i v\cdot D h_v + \bar h_v i \not\!\!D_{\bot}
   \frac{1} {(i v \cdot\! D + 2 m_Q -i\epsilon)} i \not\!\!D_\bot h_v,  \label{Lm}
 \en
${\cal L}_{Q,eff}$ is the Lagrangian of HQET, and the second term of
(\ref{Lm}) allows for a systematic expansion in terms of $i D/m_Q$.
Taking into account that ${\cal P}_+ h_v = h_v$, and using the
identity
 \be
   {\cal P}_+ i \not\!\!D_{\bot} i \not\!\!D_\bot {\cal P}_+ = {\cal P}_+
   \left[(iD_\bot)^2 + \frac{g_s}{2} \sigma_{\alpha\beta} G^{\alpha\beta}\right]{\cal P}_+,
 \en
where
 \be
   G^{\alpha\beta} = T_a G^{\alpha\beta}_a = \frac{i}{g_s}[D^\alpha,D^\beta]
 \en
is the gluon field strength tensor, thus
 \be
   {\cal L}_{Q,eff} = \bar h_v i v\cdot D h_v + \frac{1}{2 m_Q} \bar h_v
   (iD_\bot)^2 h_v + \frac{g}{4 m_Q} \bar h_v \sigma_{\alpha\beta} G^{\alpha\beta} h_v
   + {\cal O} (\frac{1}{m^2_Q}).     \label{expand}
 \en
The new operators at order $1/m_Q$ are
 \be
   {\cal O}_1 = \frac{1}{2 m_Q}~{\bar h}_v~(iD_\bot)^2~ h_v,  \label{O1} \\
  {\cal O}_2 = \frac{g_s}{4 m_Q}~{\bar h}_v~\sigma^{\mu\nu}~G_{\mu\nu}~h_v, \label{O2}
 \en
where ${\cal O}_1$ is the gauge invariant extension of the kinetic
energy arising from the off-shell residual motion of the heavy
quark, and ${\cal O}_2$ describes the color magnetic interaction of
the heavy quark spin with the gluon field. It is clear that both
${\cal O}_1$ and ${\cal O}_2$ break the flavor symmetry, while
${\cal O}_2$ breaks the spin symmetry as well. For instance, ${\cal
O}_1$ would introduce a common shift to the masses of pseudoscalar
and vector heavy mesons, and ${\cal O}_2$ is responsible for the
color hyperfine mass splittings $\delta m_{_{HF}}$.

This work did not concern the effects of strong $1/m_Q$ corrections
because they vanished when the mass difference of two ground-state
hadrons, which are the same heavy flavor but variant charge, is
taken into consideration. The full QCD Lagrangian, as $m_Q \to
\infty$, can be reduced to
 \be
  {\cal L} &=& {\cal L}_Q + {\cal L}_q + {\cal L}_g\non \\
  &\to& \bar h_v iv\cdot D h_v + \bar q~(i\gamma_\mu D^\mu - m_q)~q -
   \frac{1}{4} F^{\mu\nu}_a F_{a\mu\nu}. \label{L00}
 \en
This Lagrangian can be responsible for binding, such as a heavy
quark and a light quark in the heavy quark limit. Since an exact
solution to the QCD bound state problem does not exist, a
phenomenological approach is taken by assuming that, after summing
all the two-particle irreducible diagrams for a heavy-light system,
the effective coupling between a heavy quark ($\psi_Q$) and a light
quark ($\psi_q$) can be written as
 \be
  {\cal L}_I^{Qq} = g_0 \bar h_v i\gamma_5
  [{\cal F}(-iv\cdot \partial)\psi_q] \cdot
                   [{\cal F}(iv\cdot \partial) \bar \psi_q] i\gamma_5 h_v
 \en
in the pseudoscalar channel, where $g_0$ is a coupling constant, and
${\cal F}$ is a form factor whose presence is expected for an
effective interaction resulting from non-perturbative QCD dynamics.
${\cal L}_I^{Qq}$ can be considered as a generalized four-fermion
coupling model \cite{NJL,BH} inspired by QCD in the heavy quark
limit. If indeed the above assumption is reasonable, ${\cal
L}_I^{Qq}$ should produce a bound state of pseudoscalar heavy meson
with physical mass $m_M$. Consequently, the sum of all iterations of
diagrams should have a pole at the reduced mass
 \be
 \bar \Lambda_{\bar q} \equiv m_M-m_Q,
 \en
which is independent of the heavy flavor. However, if considering
the EM interaction, there are other contributions to $m_M$. This
will be discussed in the following section. As to the heavy baryons
which contain one (${\cal B}$) and two heavy quarks (${\cal B'}$),
we can also define the corresponding reduced masses as
 \be
 \bar \Lambda_{qq} &\equiv& m_{\cal B}-m_Q, \\
 \bar \Lambda_q &\equiv& m_{\cal B'}-m_Q-m_{Q'}.
 \en
which are independent of the heavy flavor, too.

All the above derivations concerning the Lagrangian can be suited to
the EM interaction based on the following replacements
 \be
 g_s \to e_Q e,~~ T_a \to 1,~~ A^{a\mu} \to A^\mu.
 \en
In addition, as mentioned in Section I, the EM hyperfine splitting
contribute to systematic uncertainty of the experimental results,
and can be ignored when the mass difference of heavy hadrons is
considered. We may assume that the contributions of ${\cal O}_1$ and
${\cal O}_2$ are the same order, and both are neglected here.
Then the full QED Lagrangian can be reduced to
 \be
  {\cal L} &=& {\cal L}_Q + {\cal L}_q + {\cal L}_\gamma\non \\
  &\to& \bar h_v iv\cdot D h_v + \bar q~(i\gamma_\mu D^\mu - m_q)~q -
   \frac{1}{4} F^{\mu\nu} F_{\mu\nu}. \label{LE00}
 \en
From (\ref{LE00}), we can easily derive the Feynman rules for this
Lagrangian
\begin{eqnarray}
\begin{picture}(65,30)(0,38)
\put(0,40.5){\line(1,0){40}} \put(20,40){\vector(1,0){2}}
\put(0,40){\line(1,0){40}} \put(19,28){$k$}
\end{picture}
    &:& ~~ \frac{i}{v\cdot k}\frac{1 + \not \! v}{2} ~~ ({\rm for~heavy~quark~propagator}), ~~~ \\
\begin{picture}(65,30)(0,38)
\put(0,40.5){\line(1,0){40}} 
\put(0,40){\line(1,0){40}} \put(19,40){\circle*{3}}
\end{picture}
    &:& ~~ {ie_Q e v^\mu} ~~ (\rm for~heavy~quark-photon~coupling).
\end{eqnarray}
Therefore, it can be inferred that the Coulomb energy between heavy
($Q$) and light ($q$) quarks
 \be
 \Delta E_{coul}=\alpha e_Q e_q \langle \frac{1}{r_{Qq}}\rangle
 \en
is independent of the heavy flavors.

\section{Analyses of heavy mesons and heavy baryons}
The simplest case is discussed first. As mention in Section I,
$SU(2)$ symmetry breaking comes from the up-down quark mass
difference, and the EM interactions which distinguish the different
charges carried by the up and down quarks. The mass of a $(Q\bar q)$
meson with EM Coulomb energy $e_Q e_{\bar q} \delta m_{Q\bar q}$ can
be written as
 \be
   M(Q\bar q) = m_Q + \bar \Lambda_{\bar q} + e_Q e_{\bar q} \delta m_{Q\bar q},\label{mesonmass}
 \en
where $q$ is the $u$ or $d$ quark and $\delta m_{Q\bar q}$ is
proportional to $\langle 1/r_{Q\bar q}\rangle$. Here, $SU(2)$
breaking of the $1/m_Q$ contributions is also ignored since they are
higher order effects. Thus,
 \be
   M(Q\bar d)- M(Q\bar u) = \delta \bar \Lambda_{\bar d- \bar u} + e_Q \delta m_{Q\bar
   q}\label{mdu}
 \en
where $\delta \bar \Lambda_{{\bar d}- \bar u} = \bar \Lambda_{\bar
d} -\bar \Lambda_{\bar u}$. From the experimental values
\cite{PDG08}, we can obtain
 \be
   D^+ - D^0 &=& \delta \bar \Lambda_{\bar d- \bar u} + \frac{2}{3}\delta m_{Q\bar q} = 4.78 \pm 0.1
   ~{\textrm{MeV}}, \\
   B^0 - B^- &=& \delta \bar \Lambda_{\bar d- \bar u} - \frac{1}{3}\delta m_{Q\bar q} = 0.37 \pm 0.24
   ~{\textrm{MeV}},
 \en
(the particle names stand for their masses) and consequently
 \be
   \delta \bar \Lambda_{\bar d- \bar u} &=& 1.84 \pm 0.16 ~{\textrm{MeV}},\label{d-u} \\
   \delta m_{Q\bar q} &=& 4.41 \pm 0.26 ~{\textrm{MeV}}. \label{mQbq}
 \en
As shown, $\delta \bar \Lambda_{\bar d- \bar u}$ is only
$1.84~{\textrm{MeV}}$ and smaller than the value of $m_d - m_u$
\cite{PDG08}. The reason is that since the $d$-quark is heavier, it
is also more tightly bound, so that part of the mass difference $m_d
-m_u$ is canceled by the larger binding energy of the $d$-quark. As
to the value of $\delta m_{Q\bar q}$, it provides a crucial test for
the phenomenological models within HQET to the Coulomb interaction
of QED. In addition, there are two kinds of $1/m_Q$ correction one
may consider in (\ref{mdu}), one is the strong hyperfine interaction
energy \cite{Rosner}
 \be
 \Delta E_{hf}^s=\textrm{const}\times |\Psi_{Q\bar q}(0)|^2
 \frac{\langle \sigma_Q \cdot \sigma_{\bar q} \rangle}{m_Q m_{\bar q}}.
 \en
Then $\delta \bar \Lambda_{\bar d- \bar u}$ will be replaced as
 \be
 \delta \bar \Lambda_{\bar d- \bar u} \to \delta \bar \Lambda_{\bar d- \bar u}
 +\textrm{const}\times |\Psi_{Q\bar q}(0)|^2
 \frac{\langle \sigma_Q \cdot \sigma_{\bar q}\rangle}{m_Q m_{\bar
 d}}\frac{m_{\bar u}-m_{\bar d}}{m_{\bar u}},
 \en
where we assume that $\Psi_{Q\bar d}(0)\simeq \Psi_{Q\bar u}(0)$.
The additional term is not only suppressed by $1/m_Q$, but also by
$m_{\bar u}-m_{\bar d}/m_{\bar u}$. The other is the EM hyperfine
$1/m_c$ corrections because the heavy quark limit for the charm
quark is not as good as the bottom one, then the terms such as
(\ref{Ehf}) must be added to (\ref{mesonmass}). The additional
parameters from the above two corrections will complicate
(\ref{mdu}), so that a phenomenological model need to be used to
handle the corrections.

Next, for a $(Qqq)$ baryon, its relevant mass can be written as
 \be
 M(Qqq) = m_Q + \bar \Lambda_{qq}+\sum_{i\neq j}e_i e_j \delta m_{ij},
 \en
where $i,j$ are heavy or light quarks. Here the parameterized factor
$e_q e_q \delta m_{qq}$ contains not only the Coulomb energy, but
also the hyperfine contribution. Then the mass differences of the
isospin multiplet are
 \be
 M(Qdd)-M(Quu)&=& \delta {\bar \Lambda}_{dd-uu}-2e_Q \delta
 m_{Qq}-\frac{1}{3}\delta m_{qq}, \label{dduu}\\
 M(Qdd)-M(Q\{ud\})&=& \delta {\bar \Lambda}_{dd-\{ud\}}-e_Q \delta
 m_{Qq}+\frac{1}{3}\delta m_{qq}. \label{ddud}
 \en
where $\{ud\}$ is the symmetry form $(ud+du)/\sqrt{2}$. As mentioned
in the case of heavy meson, the heavier the light degree of freedom,
the larger the binding energy $\varepsilon_{qq}$. If assuming that
there are three types of $\varepsilon_{qq}$, they are proportional
to the mass of light degree of freedom $m_{qq}$, $m^{-1/3}_{qq}$,
and independent of $m_{qq}$, which correspond to the Coulombic,
linear, and a square well potential of either finite or infinite
hight, respectively. For the first and third types, the reduced mass
$\bar \Lambda_{qq}\sim m_{qq}-\varepsilon_{qq}$ is easily checked
that it is proportional to $m_{qq}$. For the second type, the mass
differences $\delta \bar \Lambda_{dd-\{ud\}}$ and $\delta\bar
\Lambda_{\{ud\}-uu}$ can be rewritten as
 \be
 \delta\bar \Lambda_{dd-\{ud\}}&\sim& m_{dd}-m_{\{ud\}}+c
 \frac{m_{dd}-m_{\{ud\}}}{m_{dd}^{1/3}m_{\{ud\}}^{1/3}
 (m_{dd}^{2/3}+m_{dd}^{1/3}m_{\{ud\}}^{1/3}+m_{\{ud\}}^{2/3})},\non
 \\
 \delta\bar \Lambda_{\{ud\}-uu}&\sim& m_{\{ud\}}-m_{uu}+c
 \frac{m_{\{ud\}}-m_{uu}}{m_{uu}^{1/3}m_{\{ud\}}^{1/3}
 (m_{uu}^{2/3}+m_{uu}^{1/3}m_{\{ud\}}^{1/3}+m_{\{ud\}}^{2/3})},
 \en
where $c$ is a dimensional constant. For the typical values of
$m_{dd,\{ud\},uu}$, the equation
 \be
 \frac{{m_{uu}^{1/3}
 (m_{uu}^{2/3}+m_{uu}^{1/3}m_{\{ud\}}^{1/3}+m_{\{ud\}}^{2/3})}}{m_{dd}^{1/3}
 (m_{dd}^{2/3}+m_{dd}^{1/3}m_{\{ud\}}^{1/3}+m_{\{ud\}}^{2/3})}=1
 \en
is satisfied to $\sim 2\%$. Then, for the above three types of
$\varepsilon_{qq}$, $\delta\bar \Lambda_{qq-qq'}$ is almost
proportional to $m_{qq}-m_{qq'}$. In addition, following the similar
derivations, the above conclusion is also suitable to the cases that
$\varepsilon_{qq}$ is proportional to $m_{qq}^{n/n'}$ ($n$ and $n'$
are the non-zero integers). Therefore, we can obtain a relation
$\delta {\bar \Lambda}_{dd-uu}\cong 2\delta {\bar
\Lambda}_{dd-\{ud\}}$ by using the equation
$m_{dd}+m_{uu}=2m_{\{ud\}}$. Then (\ref{dduu}) and (\ref{ddud}) give
the mass difference relation
 \be
 2\Sigma_c^+ - (\Sigma_c^{++} + \Sigma_c^0) =
 2\Sigma_b^0 - (\Sigma_b^+ + \Sigma_b^-).
 \en
From the experimental values \cite{CDF,PDG08}, we have
 \be
 \Sigma_c^0 - \Sigma_c^{++} &=& -0.27 \pm 0.11~{\textrm{MeV}},\non \\
 \Sigma_c^0 - \Sigma_c^+ &=& 0.9 \pm 0.4~{\textrm{MeV}}, \non \\
 \Sigma_b^- - \Sigma_b^+ &=& 7.4 \pm 2.3~{\textrm{MeV}}, \non \\
 \Sigma_b^- &=& 5815.2 \pm 2.0 ~{\textrm{MeV}}, \non
 \en
and predict
 \be
 \Sigma_b^- - \Sigma_b^0 &=& 4.7 \pm 1.0~{\textrm{MeV}}, \\
 \Sigma_b^0 &=& 5810.5 \pm 2.2 ~{\textrm{MeV}}.
 \en
In addition, the relevant parameters in (\ref{dduu}) and
(\ref{ddud}) are obtained
 \be
 \delta \bar \Lambda_{dd-\{ud\}} &=& 2.8 \pm 0.8~{\textrm{MeV}}, \label{dd-ud}\\
 \delta m_{Qq} &=& 3.8 \pm 1.2~{\textrm{MeV}}, \label{mQq}\\
 \delta m_{qq} &=& 2.1 \pm 0.8~{\textrm{MeV}}.
 \en
Comparing (\ref{dd-ud}) and (\ref{mQq}) with (\ref{d-u}) and
(\ref{mQbq}), it is found that, for the central values, $\delta \bar
\Lambda_{dd-\{ud\}}>\delta \bar \Lambda_{\bar d- \bar u}$ and
$\delta m_{Qq}<\delta m_{Q\bar q}$. The reason is that since the
strength of the strong coupling between two quarks is smaller than
that between a quark and an antiquark, not only the canceled part of
mass difference $m_d-m_u$ in the baryon is smaller than that in the
meson, but also the expectation value of the inverse distance.
Therefore, the results lead to the above inequalities. As to a
$(Qsq)$ heavy baryon which contains one strange quark, the
corresponding mass equation is
 \be
 M(Qsq)=m_Q+\bar \Lambda_{sq}+\sum_{i\neq j}e_i e_j \delta m_{ij}.
 \en
Following a similar procedure, we can obtain:
 \be
 \Xi_b^- - \Xi_b^0 = \Xi_c^0 - \Xi_c^{+} + \delta m_{Qq}.
 \en
From the experimental data $\Xi_c^0 - \Xi_c^{+} = 3.1 \pm 0.5$ MeV
\cite{PDG08}, $\Xi_b^-=5792.9 \pm 3.0$ MeV \cite{CDF1}, and
(\ref{mQq}), we obtain the predictions
 \be
 \Xi_b^- - \Xi_b^0 &=& 6.9\pm 1.1~{\textrm{MeV}}, \\
 \Xi_b^0 &=& 5786.0 \pm 3.2~{\textrm{MeV}}.
 \en

Finally, we consider a $(QQ'q)$ doubly heavy baryon, and write its
mass as
 \be
 M(QQ'q)=m_Q+m_{Q'}+\bar \Lambda_q+\sum_{i\neq j}e_i e_j \delta
 m_{ij}.
 \en
For the two heavy quarks $(Q,Q')$ are $(c,c)$, $(b,c)$, and $(b,b)$,
we have the following results
 \be
 \Xi_{cc}^+ - \Xi_{cc}^{++}&=&\delta \bar
 \Lambda_{d-u}-\frac{4}{3}\delta m_{Qq}, \label{cc}\\
 \Xi_{bc}^0 - \Xi_{bc}^{+} &=& \delta \bar
 \Lambda_{d-u}-\frac{1}{3}\delta m_{Qq},\label{bc}\\
 \Xi_{bb}^- - \Xi_{bb}^{0} &=& (\Xi_{cc}^+ - \Xi_{cc}^{++}) +2 \delta
 m_{Qq}, \label{bb}
 \en
and the mass difference relation
 \be
 (\Xi_{cc}^+ - \Xi_{cc}^{++}) + (\Xi_{bb}^- - \Xi_{bb}^{0}) =2(\Xi_{bc}^0 - \Xi_{bc}^{+}).
 \en
The assumption $\delta \bar \Lambda_{d-u}=\delta \bar
\Lambda_{dd-\{ud\}}$ can be used because these situations are in the
baryons. From the experimental data $\Xi_{cc}^+=3518.7 \pm 1.7$ MeV
\cite{SELEX2} and the values of (\ref{dd-ud}) and (\ref{mQq}), we
can predict
 \be
 \Xi_{cc}^+-\Xi_{cc}^{++}&=& -2.3 \pm 1.7~{\textrm{MeV}}, \label{ccdu}\\
 \Xi_{cc}^{++} &=& 3521.0 \pm 2.4~{\textrm{MeV}}, \\
 \Xi_{bc}^0-\Xi_{bc}^{+}&=& 1.5 \pm 0.9~{\textrm{MeV}}, \\
 \Xi_{bb}^--\Xi_{bb}^{0}&=& 5.3 \pm 1.1~{\textrm{MeV}}.
 \en
It is worth noting that the SELEX Collaboration seeks the particle
$\Xi_{cc}^{++}$ in the corresponding decay modes \cite{SELEX3}. It
is expected that the oncoming data can confirm our calculations. In
addition, although $m_d > m_u$, the mass of $\Xi_{cc}^+(ccd)$ is
smaller than that of $\Xi_{cc}^{++}(ccu)$ from (\ref{ccdu}). The
reason is similar to the case of mass difference between
$\Sigma_{c}^+(cud)$ and $\Sigma_{c}^{++}(cuu)$, namely, since the
charge of $d$-quark is negative, the Coulomb energies between $c(u)$
and $d$ quarks reduce the masses of $\Xi_{cc}^+(ccd)$ and
$\Sigma_{c}^+(cud)$. The situations are opposite in the particles
$\Xi_{cc}^{++}(ccu)$ and $\Sigma_{c}^{++}(cuu)$. Therefore, the mass
inequalities are reversed. The predictions of this work are
summarized, and the other theoretical calculations and the
experimental data are listed in TABLE I. In previous literature,
\cite{DBL} parameterized the intrinsic quark-mass difference and the
Coulomb and magnetic-moment interactions, \cite{MIT} used the MIT
bag model, \cite{SC} studied the relativized quark model, and
\cite{BCN,SSB} used the potential models.
 \begin{table}
 \caption{\label{tab2} Experimental data, the predictions of this
 work and the other theoretical calculations (in units of MeV).}
 \begin{ruledtabular}
 \begin{tabular}{c|ccccccc}
 & Experiment & This work  & \cite{DBL} & \cite{MIT} & \cite{SC} & \cite{BCN} & \cite{SSB}  \\ \hline
 $\Sigma_c^0 - \Sigma_c^{++}$ & $-0.27\pm 0.11$ & input & $-3.4$ & $0.01$ & $-1.4$& $-0.12$ & $-1.20$  \\
 $\Sigma_c^0 - \Sigma_c^+$ & $0.9\pm 0.4$ & input & $-0.8$ & $0.83$ & $0.2$ & $0.96$ & $0.36$ \\
 $\Sigma_b^- - \Sigma_b^+$ & $7.4\pm 2.3$ & input  & & & $5.6$ & $3.58$ & $3.57$  \\
 $\Sigma_b^- - \Sigma_b^0$ &  & $4.7\pm 1.0$ & & & $3.7$ & $2.85$& $2.51$\\
 $\Xi_c^0 - \Xi_c^{+}$ & $3.1\pm 0.5$ & input & $-0.6$ & $1.72$ & & $4.67$ & $2.83$ \\
 $\Xi_b^- - \Xi_b^0$ &  & $6.9\pm 1.1$ & & & & $7.25$ & $5.39$ \\
 $\Xi_{cc}^+-\Xi_{cc}^{++}$ & & $-2.3\pm 1.7$& $-4.7$ & $-1.11$ & & $-1.87$ & $-2.96$\\
 $\Xi_{bc}^0-\Xi_{bc}^{+}$ & & $1.5\pm 0.9$ & & & & &\\
 $\Xi_{bb}^--\Xi_{bb}^{0}$ & & $5.3\pm 1.1$ & & & & &\\
 $2\Sigma_c^+-(\Sigma_c^{++}+\Sigma_c^0)$ & $-2.0\pm 0.8$ & input &  & & & $-2.04$ & $-1.92$ \\
 $2\Sigma_b^0-(\Sigma_b^++\Sigma_b^-)$ & & $-2.0\pm 0.8$ & & & & $-1.12$
 & $-0.45$
 \\ \hline
 $\Sigma_b^0$ & & $5810.5\pm 2.2$ & & & &  &\\
 $\Xi_b^0$ & &$5786.0\pm 3.2$ &   &  &  &  &\\
 $\Xi^{++}_{cc}$ & & $3521.0 \pm 2.4$&  &  &  & &
 \end{tabular}
 \end{ruledtabular}
 \end{table}
\section{Conclusions}
This study calculated the isospin mass splittings of heavy baryons
by ignoring the EM hyperfine interactions. Both the light degrees
freedom and Coulomb energies of the heavy baryons are parameterized.
In addition to deriving two mass difference equations: $2\Sigma_c^+
- (\Sigma_c^{++} + \Sigma_c^0) = 2\Sigma_b^0 - (\Sigma_b^+ +
\Sigma_b^-)$ and $(\Xi_{cc}^+ - \Xi_{cc}^{++}) + (\Xi_{bb}^- -
\Xi_{bb}^{0}) =2(\Xi_{bc}^0 - \Xi_{bc}^{+})$, we also obtained the
numerical values of some isospin mass differences. Moreover, the
masses of particles $\Sigma_b^0$, $\Xi^0_b$, and $\Xi_{cc}^{++}$ are
predicted based on the known experimental data. According to the
estimations, the decay modes $\Xi_{cc}^{++}\to \Lambda_c^+
K^-\pi^+\pi^+$ and $\Xi_{cc}^{++}\to p D^+ K^-\pi^+$ which mentioned
by the experimentalists \cite{SELEX3} are allowed. However, the
phase space of the former is obviously larger than that of the
latter.

{\bf Acknowledgments}\\
 This work is supported in part by the National Science Council of R.O.C. under Grant No:
 NSC-96-2112-M-017-002-MY3.


\end{document}